\documentclass[twocolumn,pra,preprintnumbers,amsmath,amssymb,superscriptaddress,floatfix]{revtex4}

\usepackage{latexsym}
\usepackage{epsfig}
\usepackage{graphicx}
\usepackage{textcomp}
\usepackage{color}
\usepackage{mathtools}
\usepackage{lipsum}

\begin{document}
	
\title{Single-Shot Non-Gaussian Measurements for Optical Phase Estimation}
		
\author{M. T. DiMario}
\affiliation{Center for Quantum Information and Control, Department of Physics and Astronomy, University of New Mexico, Albuquerque, New Mexico 87131, USA}

\author{F. E. Becerra}
\affiliation{Center for Quantum Information and Control, Department of Physics and Astronomy, University of New Mexico, Albuquerque, New Mexico 87131, USA}
\email{fbecerra@unm.edu}

\begin{abstract}
Estimation of the properties of a physical system with minimal uncertainty is a central task in quantum metrology. Optical phase estimation is at the center of many metrological tasks where the value of a physical parameter is mapped to the phase of an electromagnetic field and single-shot measurements of this phase are necessary. While there are measurements able to estimate the phase of light in a single shot with small uncertainties, demonstrations of near-optimal single-shot measurements for an unknown phase of a coherent state remain elusive. Here, we propose and demonstrate strategies for single-shot measurements for \textit{ab initio} phase estimation of coherent states that surpass the sensitivity limit of heterodyne measurement and approach the Cramer-Rao lower bound for coherent states. These single-shot estimation strategies are based on real-time optimization of coherent displacement operations, single photon counting with photon number resolution, and fast feedback. We show that our demonstration of these optimized estimation strategies surpasses the heterodyne limit for a wide range of optical powers without correcting for detection efficiency with a moderate number of adaptive measurement steps. This is, to our knowledge, the most sensitive single-shot measurement of an unknown phase encoded in optical coherent states.
\end{abstract}

\maketitle

The realization of measurements for precise estimation of physical quantities is essential in physics and engineering. The fundamental limits in precision for estimating a physical parameter depend on the state used to probe the system, and the process in which the parameter is encoded in the probe \cite{xu19, braun18, giovannetti13}. A central problem in quantum metrology is the determination of the fundamental limits on the achievable precision and their practical attainability \cite{braunstein94, dobrzanski15, genoni11, lee19, bradshaw18, polino20}. Achieving such quantum measurement limits given a set of physical states using physically realizable measurements is the goal of practical estimation problems.

Optical phase estimation, where information is encoded in the phase of an electromagnetic field, is essential in tasks ranging from interferometry \cite{dobrzanski15} to waveform \cite{tsang11} and force sensing \cite{iwasawa13, tsang13}. Enhanced phase estimation with and without quantum states of light has been widely investigated for sensing small deviations from a known phase \cite{anisimov10, huang17, anderson17, anderson17b, izumi16, slussarenko17}, for phase estimation with repeated sampling \cite{daryanoosh18, higgins07}, and with feedback measurements \cite{huang17, hentschel10, hou19, larson17, lumino18, zheng19}. For these particular estimation tasks, near-optimal sensitivity for phase estimation has been approached \cite{berry01, dobrzanski12, dobrzanski15}.

A different and challenging problem in parameter estimation is the realization of single-shot measurements of an unknown phase. In this estimation task, a measurement is realized to estimate the phase carried by a single optical mode in a single shot \cite{wiseman95}, either with coherent fields \cite{wiseman97, wiseman98} or quantum states of light \cite{berni15}. Single-shot measurements of a completely unknown phase of a coherent state are essential for the cooling of mechanical oscillators \cite{vanner13, seidelin19} and the preparation of spin squeezed states based on measurement backaction \cite{bouchoule02}, as well as for high-sensitivity waveform \cite{aasi13} and force detection \cite{tsang12}.

Adaptive Gaussian measurements based on homodyne detection have been extensively investigated for single-shot phase estimation with coherent states \cite{wiseman95, dariano96, wiseman97,  wiseman98, bargatin05}. These schemes can in principle outperform the heterodyne measurement limit and asymptotically approach the ultimate sensitivity for optical phase estimation given by the Cramer-Rao lower bound (CRLB) \cite{wiseman95, wiseman97, wiseman98}. Proof-of-principle experiments have demonstrated the potential of dyne adaptive measurements to surpass the heterodyne measurement limit for single-shot phase estimation of coherent states \cite{armen02} and microwave-photon wave packets \cite{martin19} after correction for experimental inefficiencies.

Here, we demonstrate optimized adaptive non-Gaussian measurements based on photon counting for single-shot phase estimation of an unknown phase of optical coherent states. These measurements use optimized coherent displacement operations, photon number resolving (PNR) detection, and conditional feedback to enable estimation strategies with high sensitivities while being robust to noise. These optimized estimation strategies allow for surpassing the heterodyne measurement limit and approaching the CRLB for coherent states with current technologies. Our experimental demonstration uses PNR detection with finite number resolution and real-time optimization of the displacement operations conditioned on the detection history as the measurement progresses. Our demonstration surpasses the heterodyne measurement limit without correcting for system inefficiencies and approaches the CRLB for coherent states when compared to a system of the same efficiency. We believe this is the most sensitive single-shot measurement of an unknown optical phase of a coherent state to date.
\begin{figure*}[t]
	\includegraphics[width = \textwidth]{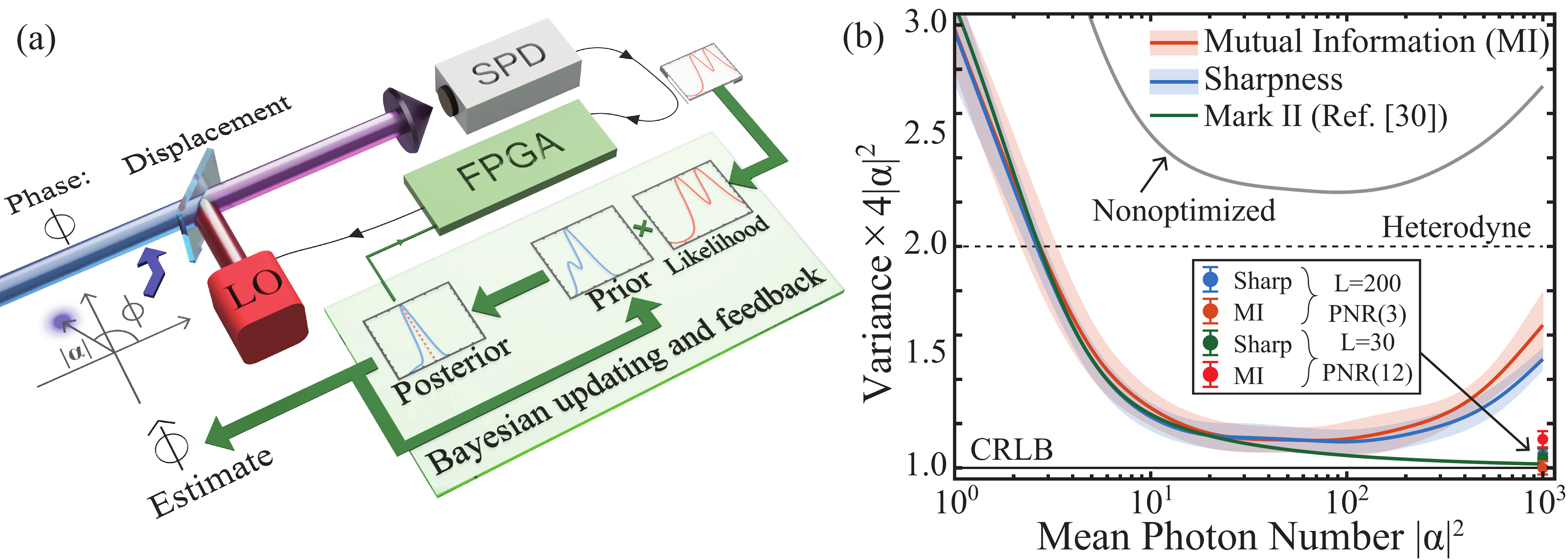}
	\caption{Optimized non-Gaussian estimation strategy. (a) Concept of the adaptive displaced photon counting measurement for single-shot phase estimation. (b) Simulated Holevo variance multiplied by the quantum Fisher information for coherent states $(4|\alpha|^2)$. Shown are strategies maximizing the sharpness $\langle S(\beta, m) \rangle$ (blue) and the mutual information $I(\beta, m)$ (orange), and a nonoptimized strategy (gray), all with $L=30$ adaptive steps, PNR(3), and ideal efficiency. Also shown is the CRLB for coherent states $(1/4|\alpha|^2)$, the lower bound of a heterodyne measurement $(1/2|\alpha|^2)$, and the performance of the adaptive homodyne scheme termed ``Mark II" from Ref. \cite{wiseman98}. Bold lines are the average of five Monte-Carlo simulations of $10^{3}$ randomly distributed initial phases, and the shaded regions represent one standard deviation. The performance of both estimation strategies at $|\alpha|^2=10^3$ for PNR(3) with $L=200$ and for PNR(12) with $L=30$ are shown for reference.
}
	\label{concept}
\end{figure*}
\newline
\newline
\emph{Non-Gaussian phase estimation strategy.---}
Figure \ref{concept}(a) shows the concept of the non-Gaussian strategy for single-shot phase estimation of a coherent state pulse in a single mode $|\alpha_{0} \rangle = |\alpha e^{i\phi_{0}}\rangle$ with a known mean photon number $\langle \hat{n} \rangle = |\alpha|^{2}$ but an unknown phase $\phi_{0}\in[0,2\pi)$. In these strategies, the displacement field is optimized in real time based on photon counting measurements within the single optical mode.

For a coherent state with unknown phase, the estimation strategy implements $L$ adaptive measurement steps over this single mode. In the first adaptive step, the input state $|\alpha_{0} \rangle$, with prior probability distribution for the phase $\mathrm{P}(\phi)=1/2\pi$, is displaced in phase space by $\hat{D}(\beta)$ using interference on a highly transmissive beam splitter to the state $\hat{D}(\beta)|\alpha_{0} \rangle = |\alpha_{0} + \beta \rangle$. The photons in the displaced state are then detected by a single photon detector described by the operators $\hat{\Pi}_{n} = | n \rangle \langle n |$ for detection of $n$ photons and $\hat{\Pi}_{m} = \hat{I} - \sum_{n=0}^{m-1}| n \rangle \langle n |$ for detection of $m$ or more photons. Here $m$ refers to the photon number resolution PNR($m$) of the detector \cite{becerra15, dimario18b}. Given the displacement field $\beta$ and detection result $n$, the strategy obtains the posterior probability distribution $\mathrm{P}(\phi|n, \beta)$ for the phase through Bayes' rule: $\mathrm{P}(\phi|n, \beta) \propto \mathcal{L}(n|\phi, \beta)\mathrm{P}(\phi)$. Here $\mathcal{L}(n|\phi, \beta)$ is the likelihood function for detecting $n$ photons given the displacement $\hat{D}(\beta)$:
\begin{equation}
\mathcal{L}(n|\phi, \beta) = \mathrm{Tr} [ \hat{\Pi}_{n} \hat{D}(\beta)|\alpha \rangle \langle \alpha | \hat{D}^{\dagger}(\beta) ] = |\langle n|\alpha + \beta \rangle|^{2}
\end{equation}
The estimate of the phase $\phi_{0}$ for this adaptive measurement step corresponds to the phase with the maximum posterior probability $\max_{\phi}\{\mathrm{P}(\phi|n, \beta)\}$. After this measurement step, the posterior distribution $\mathrm{P}(\phi|n, \beta)$ becomes the prior distribution $\mathrm{P}(\phi)$ for the subsequent adaptive measurement step, and this procedure is repeated for all $L$ adaptive periods.

Optimization of the strategy for phase estimation requires the optimization of the displacement operation $\hat{D}(\beta_{\mathrm{opt}})$ in phase $\arg(\beta)$ and amplitude $|\beta|$ in each adaptive measurement step prior to photon detection. This optimization is achieved by maximizing a given objective function that depends on the prior $\mathrm{P}(\phi)$ and the likelihood $\mathcal{L}(n | \phi, \beta)$ distributions, averaged over all possible detection results in the measurement step [see Sec. I of the Supplemental Material (SM)]. While several objective functions can be used for this optimization \cite{rodriguezgarcia20}, here we focus on two objective functions: the expected sharpness of the posterior distribution $\langle S(\beta, m) \rangle$ \cite{berry00, hentschel10, huang17} and the mutual information $I(\beta, m)$ \cite{bargatin05, rzadkowski17, paninski05}.

The expected sharpness of the posterior for a given measurement step with PNR$(m)$ is
\begin{equation}
\langle S(\beta, m) \rangle = \sum_{n=0}^{m} \mathrm{P}(n) \Bigg| \int\limits_{0}^{2\pi} e^{i\phi} \mathrm{P}(\phi | n, \beta)d\phi \Bigg| ,
\end{equation}
where P$(n) = \int \mathcal{L}(n|\phi, \beta)\mathrm{P}(\phi)d\phi$ is the probability of detecting $n$ photons. Therefore, the optimal displacement field $\beta_{\mathrm{opt}}$ maximizes $\langle S(\beta, m) \rangle$ over all possible detection results $n$ in that step, resulting in maximal expected sharpness.

The expected mutual information for an adaptive measurement step can be written as \cite{bargatin05, rzadkowski17,paninski05}
\begin{equation}
I(\beta, m) = \sum_{n=0}^{m}\int\limits_{0}^{2\pi} \mathrm{P}(\phi, n|\beta)\log_{2}\Bigg[\frac{\mathrm{P}(\phi, n|\beta)}{\mathrm{P}(n)\mathrm{P}(\phi)}\Bigg]d\phi .
\end{equation}
Here P$(\phi,n|\beta)=\mathrm{P}(\phi | n, \beta)\mathrm{P}(n) = \mathrm{P}(n | \phi, \beta)\mathrm{P}(\phi)$ is the joint probability distribution for $n$ and $\phi$ given $\beta$, and the optimal displacement $\beta_{\mathrm{opt}}$ maximizes the mutual information over all possible detection results $n$.

Recursive application of this procedure over the $L$ adaptive steps over the single-shot measurement yields a complete history of photon detections $\{n\}_{L}$ and optimal displacements $\{\beta\}_{L}$, which are used to calculate the final phase estimate:
\begin{equation}
\hat{\phi} = \mathrm{arg}\big(\langle e^{i\phi} \rangle \big) = \mathrm{arg}\Big( \int_{0}^{2\pi} e^{i\phi}\mathrm{P}(\phi|\{n\}_{L}, \{\beta\}_{L}) d\phi\Big).
\label{estim}
\end{equation}
here P$(\phi|\{n\}_{L}, \{\beta\}_{L})$ is the final reconstructed posterior distribution given the complete measurement history. After many ($N$) repetitions of the measurement with initially random relative phases between the input state and the initial displacement field, the variance of the distribution of phase estimates from Eq. (\ref{estim}) can be described by the Holevo variance for cyclic variables \cite{wiseman97, berry09, holevo11}, which is bounded by the CRLB:
\begin{equation}
\textrm{Var}[\hat{\phi}] = \frac{1}{|\langle e^{i\hat{\phi}} \rangle|^{2}} - 1 \geq \frac{1}{4|\alpha|^{2}},
\label{varh}
\end{equation}
where $| \langle e^{i\hat{\phi}} \rangle | = |\sum_{j=1}^{N}e^{i\hat{\phi}_{j}}|/N$ is the sharpness of the distribution of final estimates $\{\hat{\phi}\}_{N}$.

Figure \ref{concept}(b) shows the Holevo variance multiplied by the quantum Fisher information ($\mathrm{QFI} = 4|\alpha|^{2}$) for the optimized non-Gaussian estimation strategies with $L = 30$, PNR($3$), and ideal detection efficiency. These results are obtained through Monte Carlo simulations using different objective functions: sharpness (blue) and mutual information (orange). Bold lines represent the average of five Monte Carlo simulations each with $N=10^{3}$ randomly sampled initial phases $\phi_{0}$, and the shaded regions represent one standard deviation. Also shown is the CRLB (solid black) for coherent states given by $1/4|\alpha|^{2}$ and the lower bound on the variance of an ideal heterodyne measurement (dashed black) given by $1/2|\alpha|^{2}$.

The gray line shows the expected variance of a strategy without an optimized displacement field, which highlights the advantages of optimized strategies. A nonoptimized strategy uses a fixed local oscillator (LO) amplitude $|\beta| = |\alpha|$, but with a LO phase that is adaptively set to the current phase estimate $\hat{\phi}$. In such a strategy, the LO is always attempting to displace the estimated state $| \alpha e^{i\hat{\phi}}\rangle$ to the vacuum state. Also shown for reference is the expected performance of the adaptive homodyne strategy, termed ``Mark II'' proposed in  \cite{wiseman97, wiseman98}(green line). We observe that both optimized non-Gaussian estimation strategies enable estimation with variances below the heterodyne limit from $|\alpha|^{2} \approx 2.5 $ to $>10^{3} $ and reach a minimum of 1.13 times the CRLB at $|\alpha|^{2} \approx 50$.

We note that the increase in variance of the estimation strategies at large $|\alpha|^{2}$ can be mitigated by increasing the number of adaptive steps $L$ or the number resolution of the detector, as shown in Fig \ref{concept}(b), which we have further investigated \cite{rodriguezgarcia20}. We observe that at $|\alpha|^{2} = 10^{3}$ the strategies can reach variances within $10 \%$ from the CRLB. In particular, the strategy optimizing mutual information with $L$=200 and PNR(3) achieves the smallest variance of $1.003\pm0.03$ times the CRLB. \footnote{We note that some data sets for this point lay below the CRLB but this effect is due to finite sampling $(N=10^{3})$ in the simulations and larger $N$ would reduce this effect.} These studies suggest that optimized photon counting measurements can achieve variances within $1\%$ from the CRLB at finite mean photon numbers. Figure 1(b) also shows that for both Gaussian and non-Gaussian measurement strategies, as $|\alpha|^{2}$ decreases there is an increase in the phase variance. This behavior is consistent with that of the canonical phase measurement \cite{holevo11,wiseman98,rodriguezgarcia20}.

\begin{figure}[!tb]
	\includegraphics[width = 8.5cm]{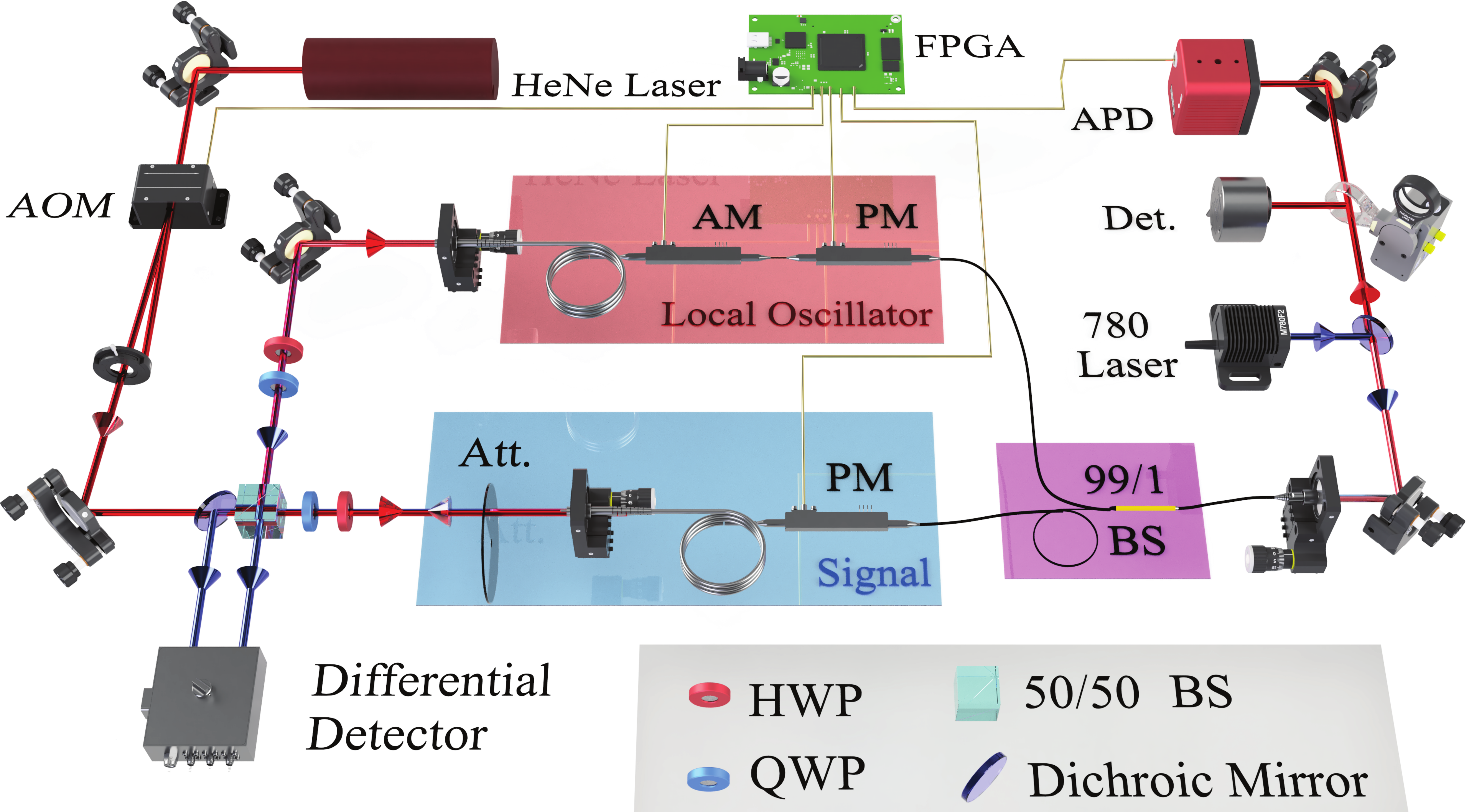}
	\caption{Experimental setup for the adaptive displaced photon counting measurement (see text for details). HeNe: helium-neon laser, AOM: acousto-optic modulator, Att.: attenuator, PM: phase modulator, AM: amplitude modulator, BS: beam splitter, APD: avalanche photodiode; FPGA: field-programmable gate array.}
	\label{setup}
\end{figure}

We find that during the initial steps of the measurement (small $k$), the optimal LO values for each objective function are in general very different due to the non-Gaussian photon counting statistics. However, as the measurement progresses, for large $k$ the posterior phase distributions for both strategies approach a Gaussian distribution with small variance, which has two notable consequences. First, in this limit the optimal LO parameters for both strategies asymptote to the same values, and these values make the classical Fisher information for displaced photon counting equal to the QFI (see Sec. II of the SM for details). This result is consistent with the theoretical work in \cite{paninski05} showing that Bayesian experimental designs that adaptively optimize the mutual information (or the variance of the posterior distribution and by extension the sharpness) are asymptotically efficient, thus reaching the CRLB \cite{cover06}. Second, in this limit of phase distributions with small variances, maximization of the mutual information is analogous to maximization of the Fisher information due to the connection between the mutual information and Fisher information via the relative entropy, i.e., Kullback-Leibler divergence \cite{kullback68,brunel98}. In contrast, maximization of the expected sharpness will yield minimal variance \cite{berry09}. This observation can be interpreted as each strategy attempting to reach the CRLB but through different approaches: either maximizing the Fisher information or minimizing the phase variance. These asymptotic findings are consistent with our results for both strategies, which show similar performances in the overall variance.
\newline
\newline
\emph{Experimental demonstration.---}
Figure \ref{setup} shows the diagram for the experimental demonstration of the optimized estimation strategies based on an interferometric setup. A helium-neon laser and an acousto-optic modulator pulsed at 1 kHz prepare coherent state pulses that enter the interferometer in a 50/50 beam splitter. An attenuator and a phase modulator prepare the input coherent state with a fixed mean photon number $|\alpha|^{2}$ and a phase $\phi_{0}$ to be estimated. The optimized displacement $\beta_{\mathrm{opt}} $ is prepared in an LO field with an amplitude modulator and a second phase modulator. We use a 99/1 beam splitter to implement the displacement operation $\hat{D}(\beta_{\mathrm{opt}})|\alpha \rangle$. The photons in the displaced state are detected by an avalanche photodiode that acts as a PNR detector with finite photon number resolution \cite{becerra15, dimario18b}. A field-programmable gate array (FPGA) collects the detection events and implements the phase estimation strategies described above. This FPGA realizes in real time the Bayesian estimation procedure during each adaptive measurement, as described in Sec. I of the SM, and optimizes the displacement $\hat{D}(\beta_{\mathrm{opt}})$ conditioned on photon detections. The optimization protocol in the FPGA uses a Gaussian approximation for steps $k\geq9$ in which the phase distributions $\mathrm{P}(\phi)$ are approximated as Gaussian distributions (see Sec. II in the SM). This approximation allows us to demonstrate strategies with large numbers of adaptive steps, $L=30$. At the end of an estimation measurement, the FPGA transfers the history of detections $\{n\}_{L}$, the phase estimates $\{\hat{\phi}\}_{L}$, and the optimized displacements $\{\beta_{\mathrm{opt}}\}_{L}$  to a computer for processing and final estimation based on Eq. (\ref{estim}). We note that while the optimization protocol uses a Gaussian approximation, the recursive Bayesian procedure uses the complete phase distribution $\mathrm{P}(\phi)$.

For experimental convenience, instead of randomly choosing the initial phase $\phi_{0}$ and a fixed initial displacement phase $\theta_{\beta}^{(0)}=\arg(\beta)=0$, we fix $\phi_{0} = \pi$ and start at a random phase $\theta_{\beta}^{(0)}$. These two situations are equivalent, since the initial relative phase $\theta_{\beta}^{(0)}-\phi_{0}$ is random in both cases (see Sec. III of the SM). Our experimental implementation has a $\approx66\%$ duty cycle, uses a laser at 780 nm to stabilize the interferometric setup \cite{becerra13, becerra15, ferdinand17, dimario18b}, and achieves an overall system efficiency $\eta = 0.70$ (avalanche photodiode efficiency efficiency $\eta_{APD}\approx$ 0.84), interference visibility $\xi = 0.997$, and dark count rate $\nu = 140/s$. We demonstrate estimation strategies for mean photon numbers from $|\alpha|^{2}=1$ to $|\alpha|^{2}=10^{3}$, with PNR(3) and $L$ = 30 adaptive measurement steps that are each 20 $\mu$s long.
\begin{figure}[!t]
	\includegraphics[width = 8.5cm]{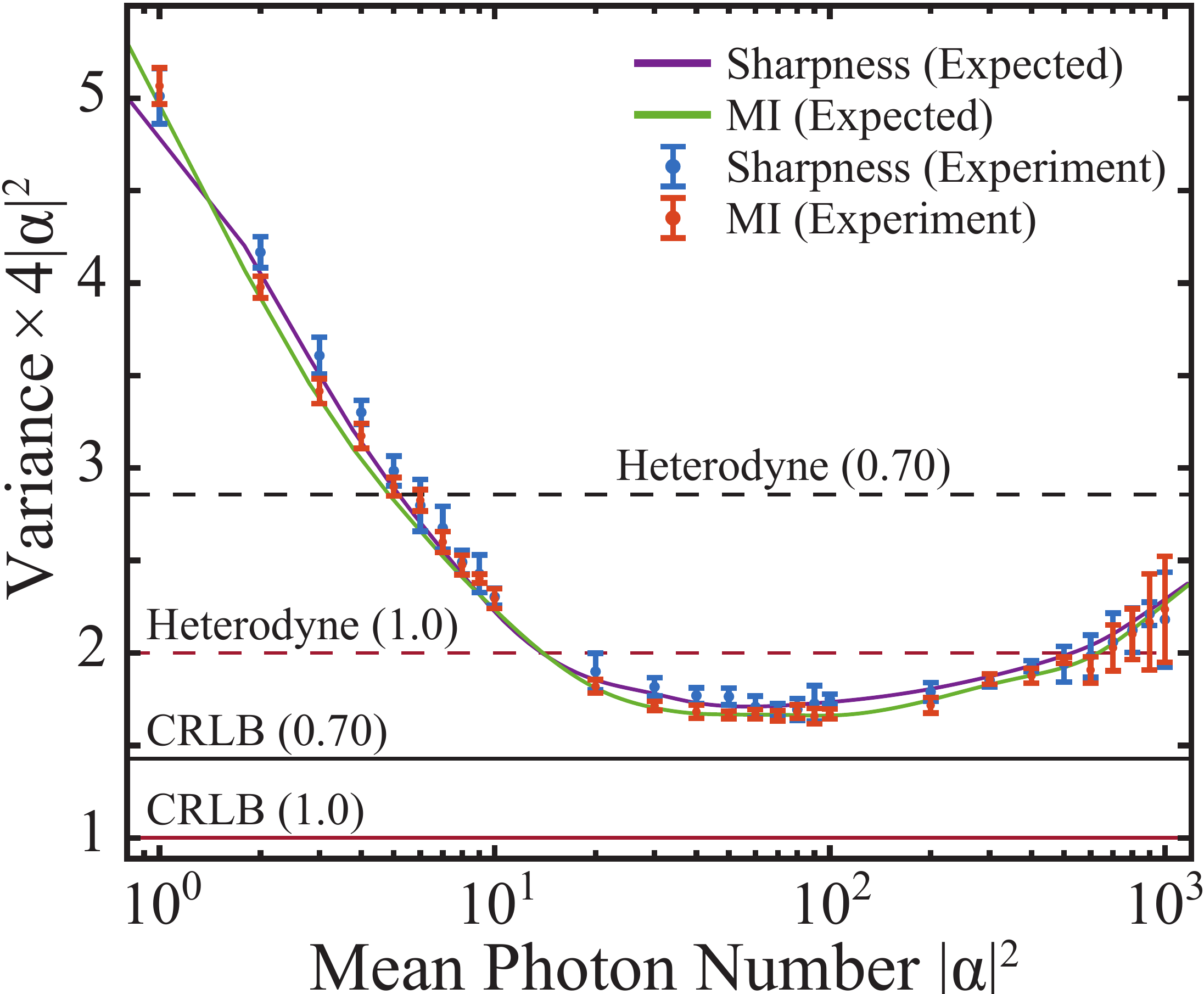}
	\caption{Experimental results. Experimentally obtained Holevo variance multiplied by the QFI for coherent states ($4|\alpha|^{2}$) for estimation strategies maximizing the sharpness and the mutual information as blue and orange points, respectively. Also shown are the expected variances (solid lines) accounting for experimental parameters. Included are the CRLB for coherent states $1/4|\alpha|^2$ (solid red line) and the ideal heterodyne limit $1/2|\alpha|^2$ (dashed red line), together with these bounds adjusted to the efficiency of our implementation $\eta=0.70$ (solid and dashed black lines). Note that both strategies outperform the ideal heterodyne bound from $|\alpha|^{2}\approx20$ to $\approx 600$ and the adjusted bound from $|\alpha|^{2}\approx6$ to $ > 10^{3}$. Furthermore, both estimation strategies achieve a minimum of $\approx$ 1.68 times the CRLB (1.18 times adjusted) at $|\alpha|^{2} \approx 70$.}
	\label{data}
\end{figure}
\newline
\newline
\emph{Results.---}Figure \ref{data} shows, as a function of $|\alpha|^{2}$, the Holevo variance of the experimental results for single-shot phase estimation of coherent states multiplied by the QFI ($4|\alpha|^{2}$). Data points are the results for strategies maximizing the average sharpness $\langle S(\beta, m) \rangle$ (blue) and the mutual information $I(\beta, m)$ (MI, orange). The points and error bars represent the average variance and one standard deviation over five different runs of the experiment, respectively, each of which is the variance of $N=10^{4}$ independent experiments with random initial relative phases $\theta_{\beta}^{(0)}-\phi_{0}$. The CRLB (solid red line), the heterodyne limit (dashed red line), and these bounds adjusted to the efficiency $\eta=0.70$ (solid and dashed black lines) are included for reference. The purple (green) solid line is the expected Holevo variance for a strategy maximizing the sharpness (mutual information) obtained from the average of five Monte Carlo simulations of the experiment, each with $10^{3}$ samples. These simulations take into account the experimental imperfections and the effects of limited resolution, precision, and bandwidth of the FPGA on the implementation of the strategies. The increase of the error bars for large $|\alpha|^{2}$ in the experimental results is due to the increased sensitivity of the strategies to environmental noise at larger mean photon numbers. As the variance is small in this regime, technical noise causing small fluctuations or instabilities in the relative phase of the input state and LO may become non-negligible compared to the measured variance.

We observe that the optimized non-Gaussian estimation strategies surpass the ideal heterodyne bound from $|\alpha|^{2}\approx 20$ to 600, and the adjusted ($\eta=0.70$) heterodyne bound from $|\alpha|^{2} \approx 6$ to $> 10^{3}$. Furthermore, in both strategies the variance reaches a minimum at $|\alpha|^{2} \approx 70$ of less than 1.7 times the CRLB, and less than 1.2 times adjusted for $\eta=0.70$. These results show that adaptive non-Gaussian measurements enable highly robust phase estimation below the heterodyne limit without correction for detection inefficiencies. We note that increasing the number of adaptive steps and photon number resolution \cite{rodriguezgarcia20} could allow for increasing the measurement sensitivity at higher $|\alpha|^{2}$.
\newline
\newline
\emph{Conclusion.---}
We propose and demonstrate optimized non-Gaussian estimation strategies for single-shot measurements of an unknown phase of optical coherent states with sensitivities surpassing the heterodyne limit and approaching the CRLB. These strategies use optimized displacement operations, single-photon counting with finite photon number resolution, and a moderate number of adaptive steps with fast feedback. Our demonstration of the single-shot estimation strategies uses fast processing for optimization of the displacement operations in real time conditioned on the detection history as the measurement progresses throughout the single optical mode. We show that our experimental demonstration surpasses the ideal heterodyne limit for a wide range of optical powers without correcting for detection efficiency and achieves variances of less than 1.2 times the equivalent CRLB adjusted to our system efficiency. This is, to our knowledge, the most sensitive single-shot measurement of an unknown phase encoded in optical coherent states to date.

We expect that these optimized measurements for phase estimation of coherent states can be used to enhance the performance of schemes based on measurement backaction for the cooling of mechanical oscillators \cite{vanner13, seidelin19}, the preparation of spin squeezed states \cite{bouchoule02}, and applications in waveform and force detection \cite{tsang12}. Moreover, optimized non-Gaussian measurements can potentially be used for surpassing the limits of Gaussian approaches for estimating a time-varying phase \cite{zhang19, wheatley10, pope04, yonezawa12} and maximizing information transmission in optical communications \cite{ferdinand17, lee16}, as well as for phase estimation when quantum resources are employed \cite{berni15}.

\begin{acknowledgments}
This work was supported by the National Science Foundation (NSF) (PHY-1653670, PHY-1521016, and PHY-1630114).
\end{acknowledgments}


\end{document}


\title{Supplemental Material: Single-shot non-Gaussian Measurements for Optical Phase Estimation}

\author{M. T. DiMario}
\affiliation{Center for Quantum Information and Control, Department of Physics and Astronomy, University of New Mexico, Albuquerque, New Mexico 87131}

\author{F. E. Becerra}
\affiliation{Center for Quantum Information and Control, Department of Physics and Astronomy, University of New Mexico, Albuquerque, New Mexico 87131}
\email{fbecerra@unm.edu}
	
\begin{abstract}
		
\end{abstract}
	
\maketitle

The Supplementary Material describes in Section I the FPGA algorithm used to implement the adaptive measurement; in Section II the Gaussian prior distribution approximation used to perform the experiment and the connection with the classical Fisher information for displaced photon counting; and in section III it shows that the estimator based on adaptive non-Gaussian measurements is unbiased with respect to the initial relative phase.

\begin{figure*}[!tbp]
	\includegraphics[width = \textwidth]{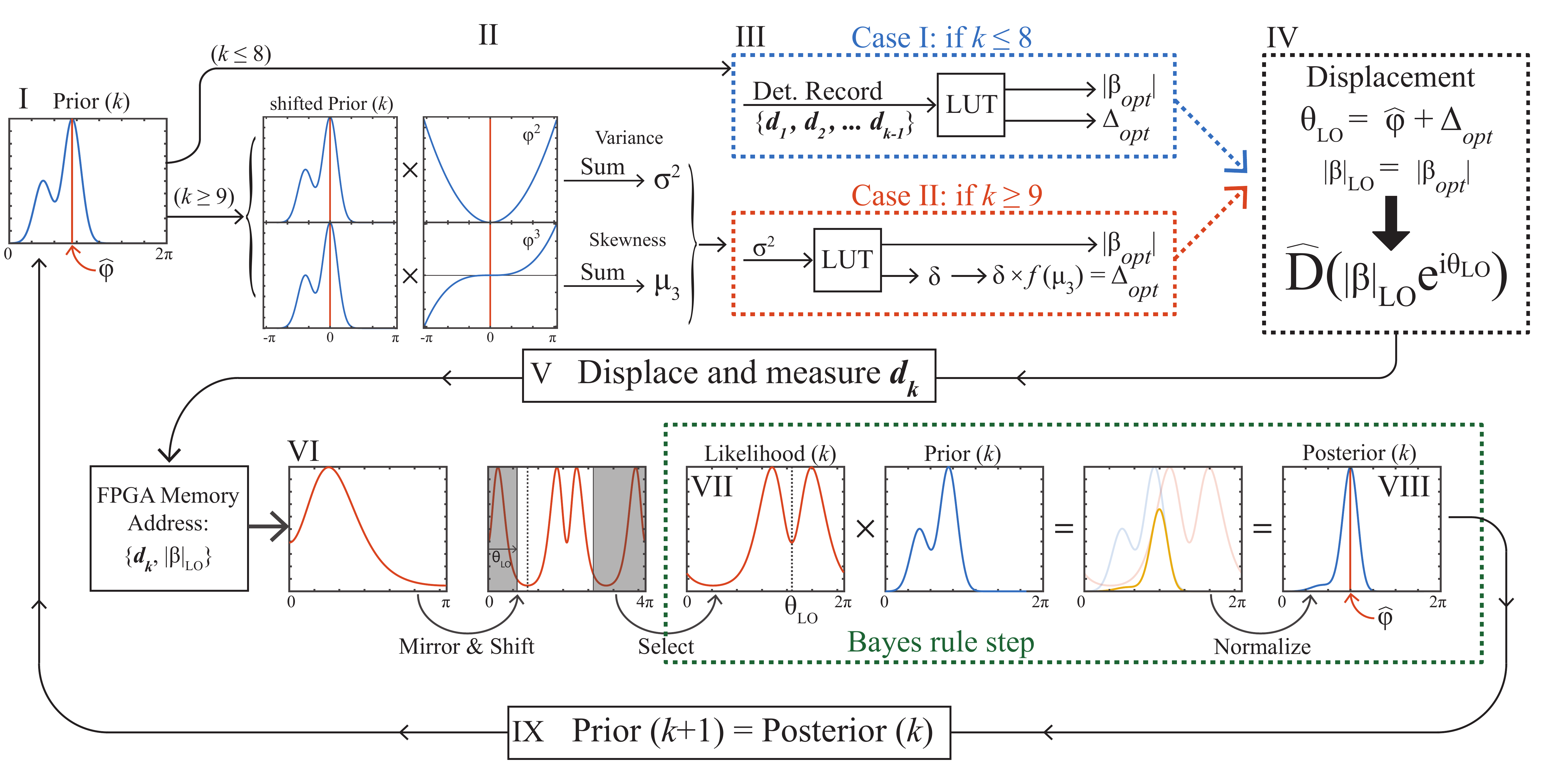}
	\caption{FPGA algorithm for the optimized estimation strategies. The flowchart shows the procedure used in the FPGA for obtaining the optimal LO amplitude $|\beta_{opt}|$ and phase $\theta_{opt}$, posterior probability distribution, and phase estimate $\hat{\phi}$ for the measurement step $k$. Based on $\mathrm{P}_{prior}(\phi)$ at adaptive step $k$, the FPGA determines $|\beta_{opt}|$ and $\Delta_{opt}$ from the full calculation for steps $k \leq 8$, or using a Gaussian approximation for $\mathrm{P}_{prior}(\phi)$ for steps $k \geq 9$. See text for details. The FPGA uses the detection result and LO amplitude for step $k$ to obtain the posterior probability phase distribution $\mathrm{P}_{post}(\phi)$ for adaptive step $k$. The estimation strategy uses a recursive Bayesian procedure where the posterior phase distribution $\mathrm{P}_{post}(\phi)$ in step $k$ is defined as the prior distribution $\mathrm{P}_{prior}(\phi)$  for step $k+1$.
	}
	\label{flowchart}
\end{figure*}

\section{Optimization Algorithm}

Figure \ref{flowchart} shows a flowchart of the phase estimation algorithm implemented in real time in the FPGA to obtain the optimal values of the displacement field amplitude $|\beta_{opt}|$ and phase $\theta_{opt}$ for each adaptive step $k$, and to perform the Bayesian updating. This algorithm has two cases depending on the current adaptive step $k$. For $k\leq8$ the algorithm requires full calculation of the optimal values $|\beta_{opt}|$ and $\theta_{opt}$ using the complete prior phase distributions. For $k\geq9$ the algorithm approximates the prior phase distributions as Gaussian distributions allowing for a very efficient estimation of $|\beta_{opt}|$ and $\theta_{opt}$ in real time in the FPGA.

Starting at I in figure \ref{flowchart}, given the prior probability distribution $\mathrm{P}_{prior}(\phi)$ at adaptive step $k$, the FPGA calculates the optimal displacement amplitude $|\beta_{opt}|$ and phase $\theta_{opt}$ for the current measurement step (shown as I-III in Fig. \ref{flowchart}). For convenience, $\theta_{opt}$ is expressed as the sum of the current phase estimate $\hat{\phi}$ and an optimal phase offset $\Delta_{opt}$, $\theta_{opt} = \hat{\phi} + \Delta_{opt}$. Two cases can be identified depending on the measurement progression in $k$:

\textbf{Case I.--} For $k\leq8$ the optimal values $|\beta_{opt}|$ and $\Delta_{opt}$ are obtained from a lookup table (LUT) whose input is the complete detection record $\{d_{1}, d_{2}, ... d_{k-1}\}$ up to the adaptive step $k$ (see dashed blue rectangle). This LUT is pre-calculated for all possible detection records based on the maximization of the expected sharpness or the mutual information to provide the exact optimal values of $|\beta_{opt}|$ and $\Delta_{opt}$ using the full probability distribution $\mathrm{P}_{prior}(\phi)$ for every adaptive step up to $k$.

\textbf{Case II.--}  If $k\geq9$ the FPGA switches to a mode where the prior distribution  $\mathrm{P}_{prior}(\phi)$ for step $k$ is approximated as Gaussian (Gaussian approximation) in order to calculate approximately optimal values $|\beta_{opt}|$ and $\Delta_{opt}$ in an efficient way. This approximation is justified by observing that after $k \geq 9$ steps, the prior distribution $\mathrm{P}_{prior}(\phi)$ is close to Gaussian, so that it can be described by its first three moments: the mean, the variance $\sigma^{2}$, and skewness $\mu_{3}$ of the distribution, while its support over higher order moments is negligible. In particular, the skewness $\mu_{3}$ mainly describes the deviation of the prior $\mathrm{P}_{prior}(\phi)$ from a Gaussian distribution and provides information to correct the inferred LO phase offset $\delta$ under this approximation to obtain $\Delta_{opt}$ in step $k$.

To obtain the optimal values $|\beta_{opt}|$ and $\Delta_{opt}$ for $k\geq9$, the FPGA calculates the variance $\sigma^{2}$ and skewness $\mu_{3}$ of $\mathrm{P}_{prior}(\phi)$ (shown as II in Fig. \ref{flowchart}). The variance $\sigma^{2}$ is used to calculate the optimal amplitude $|\beta_{opt}|$ and phase $\delta$ under the Gaussian approximation using a second LUT with pre-calculated values (see dashed orange rectangle). Then, the phase $\delta$ obtained under this approximation is corrected by a function $f(\mu_{3})$ of skewness:

\begin{equation}
\Delta_{opt} = f(\mu_{3}) \times \delta \text{ , with  } f(\mu_{3})=-\frac{\textrm{sign}(\mu_{3})}{1+|\mu_{3}|}.
\label{skew}
\end{equation}

This function provides the correction from this approximation, and is empirically found through Monte Carlo simulations by comparing the optimal phases obtained from a Gaussian distribution and the full distribution with nonzero $\mu_{3}$. Note that calculating the optimal displacement values takes approximately 500 ns.

Given obtained $|\beta_{opt}|$ and $\Delta_{opt}$, the offset $\Delta_{opt}$ is added to the current phase estimate $\hat{\phi}$ at step $k$ to define the amplitude and phase $\theta_{opt}$ of the displacement operation for step $k$ (see dashed black rectangle). This displacement is implemented and a photon detection result $d_{k}$ for step $k$ is acquired (shown as V in Fig. \ref{flowchart}).

To obtain $\mathrm{P}_{post}(\phi)$ for step $k$, the FPGA obtains the likelihood function $\mathcal{L}(d_{k}|\phi, \beta)$ for the detected number of photons. Based on the photon detection result $d_{k}$ and the amplitude $|\beta|$ of local oscillator field, the FPGA retrieves from memory a segment of the likelihood function corresponding to the phase range of $\phi\in[0,\pi)$ (shown as VI in Fig. \ref{flowchart}). The FPGA then constructs the complete likelihood function using the fact that the likelihoods are symmetric about $\phi=\theta_{LO}$ and  periodic with period $2\pi$ (shown as VII in Fig. \ref{flowchart}). The posterior distribution $\mathrm{P}_{post}(\phi)$ for step $k$ is then calculated through Bayes' rule as the product of the likelihood $\mathcal{L}(d_{k}|\phi, \beta)$ and the prior probability distribution $\mathrm{P}_{prior}(\phi)$ (see dashed green rectangle). We note that the Bayesian updating uses the full prior probability distribution and the Gaussian prior approximation is used only for the LO optimization when $k\geq9$.

An updated phase estimate $\hat{\phi}$ is then obtained as the phase with maximum posterior probability. Next, the FPGA performs a quasi-normalization of the posterior (shown as VIII in Fig. \ref{flowchart}) by multiplying the posterior by $2^{p}$ where $p$ is an integer such that the most significant bit of the updated phase estimate (maximum of the posterior) is one. This prevents computational underflow in the Bayesian step of the algorithm so that the FPGA can maintain an accurate distribution for the phase for the entire measurement. The FPGA uses discretized probability distributions with 256 phase points and each probability value is represented in fixed point notation as a 10 bit number. Finally, given the posterior distribution $\mathrm{P}_{post}(\phi)$ for step $k$, the recursive Bayesian algorithm now defines the posterior $\mathrm{P}_{post}(\phi)$ in step $k$ as the prior $\mathrm{P}_{prior}(\phi)$ for the next step $k+1$ (shown as IX in Fig. \ref{flowchart}). The Bayesian updating takes approximately 500 ns, giving a feedback bandwidth of $\approx 1 \mathrm{MHz}$.

We note that as the input mean photon number $|\alpha|^{2}$ increases, the posterior phase distributions in general become very narrow (i.e. have small variance). With increasing $|\alpha|^{2}$, accurately representing the phase distributions in the simulations and in the experimental implementation requires increasing the number of discretization points for the phase. A larger number of discretization points will increase the computational complexity of the algorithm roughly linearly with the number of points. Within the FPGA, the increase in complexity is mainly due to an increase in the number of multiplications needed to calculate the posterior phase distribution, which increases the computation time and the memory required to store the likelihood functions. While the resources required for the experimental implementation increase with $|\alpha|^{2}$, in theory this does not affect the complexity of the proposed strategy for phase estimation. On the other hand, if the PNR used in the strategy is increased, the time required to compute the expected sharpness in Eq. (2), and mutual information in Eq. (3), will increase linearly with the number of terms in the sums in Eq. (2) and Eq. (3).

\begin{figure*}[!tp]
	\includegraphics[width = 0.9\textwidth]{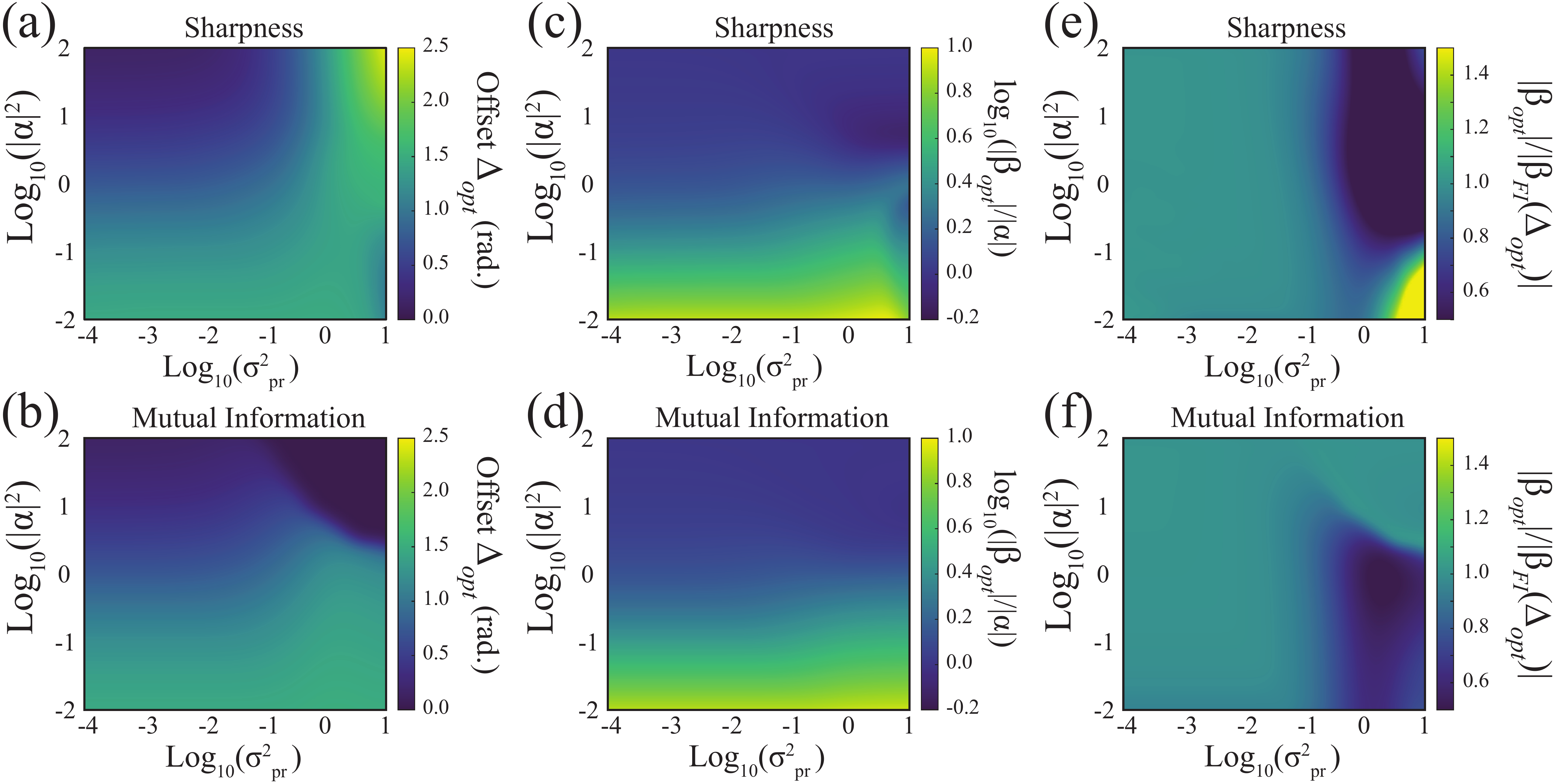}
	\caption{Optimal LO amplitude and phase based on the Gaussian approximation for Gaussian prior distributions with different mean photon numbers $|\alpha|^2$ and variances $\sigma_{pr}^{2}$. (a,b) Optimal phase offset $\Delta_{opt}$ for estimation strategies maximizing (a) the expected sharpness and (b) the mutual information in each adaptive step. (c,d) Ratio of the optimal LO amplitude $|\beta_{opt}|$ to $|\alpha|$ on a log scale for strategies maximizing (c) the sharpness and (d) the mutual information. (e,f)  Ratio of the optimal LO amplitude $|\beta_{opt}|$ to the amplitude maximizing the Fisher information $|\beta_{FI}(\Delta_{opt})|$ from Eq. (\ref{optcond_delta}) for strategies maximizing (e) the sharpness and (f) the mutual information. Note the similarity for optimal phase offsets and amplitudes for both estimation strategies. Also note that the ratio $|\beta_{opt}|/|\beta_{FI}(\Delta_{opt})|$ approaches 1 as $\sigma_{pr}^{2}$ decreases for all mean photon numbers.
}
	\label{varlut}
\end{figure*}
\section{Gaussian Approximation}
In general, accurately describing the prior phase distribution $\mathrm{P}_{prior}(\phi)$ at step $k$ requires specifying many moments of the distribution. However, as mentioned above, in the adaptive estimation strategy based on recursive Bayesian updating, for $k \geq 9$ the prior $\mathrm{P}_{prior}(\phi)$ at step $k$ can be approximately described by its first three moments: mean, variance, and skewness. We find that the prior distribution can be well approximated as a Gaussian distribution with the mean equal to the current phase estimate $\hat{\phi}$ at the adaptive step $k$ and a variance $\sigma_{pr}^{2}$ equal to the variance of the prior distribution:
\begin{equation}
\sigma_{pr}^{2} = \int_{0}^{2\pi} (\phi - \hat{\phi})^{2} \mathrm{P}_{prior}(\phi),
\label{VarPrior}
\end{equation}
so that
\begin{equation}
\mathrm{P}_{prior}(\phi)\approx\tilde{\mathrm{P}}(\phi) = \mathcal{N}e^{-\frac{(\phi - \hat{\phi})^{2}}{2\sigma_{pr}^{2}}}
\label{Gaussian}
\end{equation}
where $\mathcal{N}$ is a normalization factor.

Given the calculated variance $\sigma_{pr}^{2}$ from Eq. (\ref{VarPrior}), it is possible to very efficiently obtain optimal values of the LO field for the amplitude $|\beta_{opt}|$ and the phase $\delta$ under the Gaussian approximation by optimizing the mutual information or sharpness for a Gaussian distribution. This is possible because the Gaussian distribution is uniquely defined by its mean and its variance, see Eq. (\ref{Gaussian}). Moreover, since we are looking for an optimal phase offset, we can find the optimal values for a zero-mean Gaussian, and then shift the phase offset $\Delta_{opt}$ by $\hat{\phi}$ to obtain the optimal phase $\theta_{opt}$ for the LO field. As a result, for a given  mean photon number $|\alpha|^{2}$, it is possible to construct a look up table (LUT) in the FPGA that associates the variance $\sigma_{pr}^{2}$ to the optimal values of the LO amplitude $|\beta_{opt}|$ and phase offset $\Delta_{opt}$, after the correction with the function $f(\mu_{3})$ in Eq. (\ref{skew}) (also see Fig. \ref{flowchart}).

Figure \ref{varlut} shows the calculated optimal values for the LO amplitude $|\beta_{opt}|$ and phase offset $\Delta_{opt}$ for different mean photon numbers $|\alpha|^{2}$ as a function of the variance of a Gaussian prior distribution $\tilde{\mathrm{P}}(\phi)$. Panels (a) and (b) show $\Delta_{opt}$ for estimation strategies maximizing the sharpness and the mutual information, respectively. Panels (c) and (d) show $|\beta_{opt}|$ for the same cases, respectively. We note that for small values of $\sigma_{pr}^{2}$, the optimal amplitude and phase values for the two different objective functions asymptote to the same value, which indicates that under the Gaussian approximation both estimation strategies are expected to show similar performances for large values of $k$.
\\
\\
\emph{\textbf{Asymptotic behavior}.---}
To further investigate the expected asymptotic behavior of the estimation strategies, we study the classical Fisher information (CFI) for measurements based on displacement and photon counting, which can provide insight into their performance compared to the fundamental lower bound, the Cramer-Rao lower bound. The CFI for displaced photon counting, with infinite photon number resolution (PNR), is given by:
\begin{align}
F(\phi) &= \sum_{n=0}^{\infty}\mathcal{L}(n|\phi) \Bigg( \frac{\partial}{\partial \phi}\mathrm{ln}\big(\mathcal{L}(n|\phi) \big) \Bigg) ^{2}
\nonumber
\\
\nonumber
\\
&= \frac{4|\alpha|^{2}|\beta|^{2}\mathrm{sin}^{2}(\phi)}{|\alpha|^{2} + |\beta|^{2} - 2|\alpha| |\beta| \mathrm{\cos}(\phi)}
\label{fieq}
\end{align}

Where $\mathcal{L}(n | \phi)$ is the likelihood function for detecting $n$ photons given the relative phase $\phi = \phi_{0} - \theta_{LO}$ between the input state and the LO. Note that when $|\beta|=|\alpha|$ the CFI of photon counting measurements is equal to the quantum Fisher Information (QFI) when $\phi=0$, which corresponds to performing displacement to the vacuum state. However, for estimation of an unknown phase, it is possible to find the optimal LO amplitude $|\beta|$ that maximizes $F(\phi)$:
\begin{equation}
|\beta_{FI}(\phi)|^{2} = \frac{|\alpha|^{2}}{\mathrm{cos}^{2}(\phi)}
\label{optcond}
\end{equation}

The evaluation of $F(\phi)$ in $|\beta_{FI}(\phi)|$ results in:
\begin{equation}
FI(\phi) \bigg \rvert_{|\beta|=|\beta_{FI}(\phi)|} = 4|\alpha|^{2} = QFI
\end{equation}

The CFI for displaced photon counting is guaranteed to be equal to the QFI for coherent states as long as Eq. (\ref{optcond}) is satisfied. Therefore, for any given relative phase $\phi$, there is a $|\beta_{FI}(\phi)|$ that saturates the QFI.

For the optimized estimation strategies based on adaptive measurements, after many adaptive steps the prior probability distributions approach Gaussian distributions with small variance $\sigma_{pr}^{2}$, and the estimate $\hat{\phi}$ approaches the true phase $\phi_{0}$. In this asymptotic limit, one may expect that the CFI of the estimation strategies approaches the QFI, and the amplitude $|\beta_{opt}|$ and phase $\Delta_{opt}$ approach values that satisfy the optimality condition in Eq. (\ref{optcond}), where $\Delta_{opt}$ describes the relative phase between the displacement field $\theta_{LO}$ and the actual phase $\phi_{0}$. That is:

\begin{equation}
\lim\limits_{\sigma_{pr} \to 0}|\beta_{opt}|^{2} = |\beta_{FI}(\Delta_{opt})|^{2} = \frac{|\alpha|^{2}}{\mathrm{cos}^{2}(\Delta_{opt})}
\label{optcond_delta}
\end{equation}

Figures \ref{varlut}(e) and \ref{varlut}(f) show the ratio $|\beta_{opt}|/|\beta_{FI}(\Delta_{opt})|$ of the optimal LO amplitude to the expected optimal amplitude from Eq. (\ref{optcond_delta}) given $\Delta_{opt}$ for strategies maximizing (e) the sharpness $\langle S(\beta, m) \rangle$ and (f) the mutual information $I(\beta, m)$. We observe that as $\sigma_{pr}$ decreases, the optimal amplitudes $|\beta_{opt}|$ obtained in the Gaussian approximation asymptote to $|\beta_{FI}(\Delta_{opt})|$ for $\mathrm{log_{10}}(\sigma_{pr}^{2}) < -1.5$ for all mean photon numbers. We note that while the optimal values $|\beta_{opt}|$ and $\Delta_{opt}$ in the estimation strategies without the Gaussian approximation during small number of steps $k$ do not satisfy the condition in Eq. (\ref{optcond_delta}), the value of $|\beta_{opt}|$ approaches $|\beta_{FI}(\Delta_{opt})|$ for large $k$.

\section{Preparation of initial phase}

In an ideal experiment for estimation of a completely unknown phase $\phi_{0}$, the initial phase of the  local oscillator (LO) displacement field $\theta_{LO}$ would be fixed, and the phase to be estimated would be chosen randomly $\phi_{0}\in[0,2\pi)$. This means that the initial relative phase $\theta_{LO}-\phi_{0}$ is random. However, since in our experimental setup we had arbitrary control of the phase $\theta_{LO}$ with high bandwidth for the adaptive measurements, we chose to fix $\phi_{0}$, and have the initial LO phase $\theta_{LO}$ be uniformly distributed, so that the initial relative phase is still random. In particular, in our experiment we set $\phi_{0} = \pi$ and randomly choose the initial LO phase $\theta_{LO}\in[0,2\pi)$, which is prepared with a 8-bit resolution DAC resulting in $2^8=256$ different possible initial phases.

\begin{figure}[!bp]
	\includegraphics[width = 8.5cm]{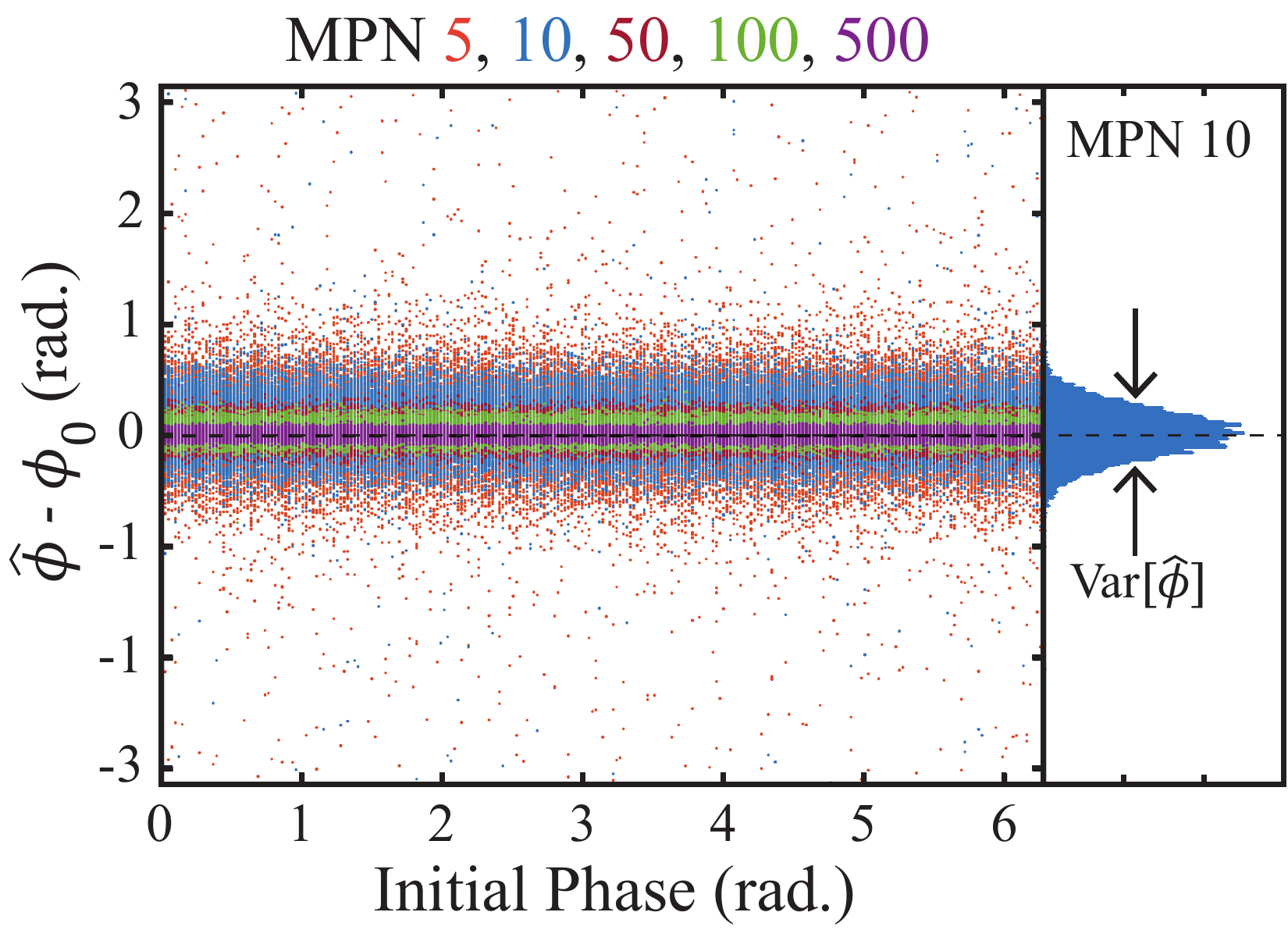}
	\caption{Estimator bias. Estimator error $\hat{\phi}-\phi_{0}$ for each initial LO phase $\theta_{LO}\in[0,2\pi)$ for every experimental run ($5\times 10^{4}$ samples) for different total mean photon numbers for a strategy maximizing the mutual information in each adaptive step. Note the independence of the error on the initial LO phase.}
	\label{error}
\end{figure}

To show that there is not any bias introduced by this procedure, we studied the phase estimator error $\hat{\phi} - \phi_{0}$ obtained from the estimation of $\phi_{0}$ with different initial LO phases $\theta_{LO}\in[0,2\pi)$. Figure \ref{error} shows the final phase error after $L=30$ adaptive steps for all experiments (total of $5\times 10^{4}$) for an estimation strategy maximizing the mutual information in each adaptive step. Each point represents one single estimation experiment, and the figure shows results for total mean photon numbers of $|\alpha|^{2}$ = 5 (orange), 10 (blue), 50 (red), 100 (green), and 500 (purple). We note the absence of any dependence of the final estimates $\hat{\phi}$ on the initial LO phase $\theta_{LO}$. Figure \ref{error} also shows an example of the distribution of final phase estimates $\hat{\phi}$ for $|\alpha|^{2} = 10$ showing normally distributed behavior. The variance Var[$\hat{\phi}$] of this distribution is approximately equal to the mean square error of the estimator for $|\alpha|^{2} = 10$.

